\documentclass[12pt]{article}
\usepackage[dvips]{graphicx}
\newcommand{\beq}{\begin{equation}}
\newcommand{\eeq}{\end{equation}}
\newcommand{\beqa}{\begin{eqnarray}}
\newcommand{\eeqa}{\end{eqnarray}}
\newcommand{\IP}{I\!\!P}
\begin{document}  
\title{Pomeron contribution to Spin-flip\thanks{
Work supported in part by the MURST of Italy}.}
\author{A.F. Martini\thanks{Fellow of FAPESP, SP, Brazil} 
and E. Predazzi \\
Depto. Fisica Teorica, Univ. of Torino - Italy \\
and \\
INFN - Sezione di Torino - Italy}
\date{}
\maketitle
\abstract{The $pp$ polarization data confirm the presence of a
diffractive-like (Pomeron) contribution in the spin flip amplitude. 
The extrapolation to RHIC energies does not appear very promising.}

\section{Introduction}

Many years ago \cite{predazzi67}, one of us noticed that earlier high 
energy polarization data suggested a diffractive contribution in the 
spin-flip $\pi p$ amplitude at high energy which becomes evident when 
the kinematical zero is removed. This contribution manifests itself 
in a ``{\it reduced}'' spin-flip amplitude which is peaked in the 
forward direction and which does not vanish when the energy increases. 
In Ref. \cite{predazzi67} it was explicitly noticed that, once the 
kinematical zero is reduced, all partial waves act coherently in the 
small angle domain as it is typical of diffractive events 
\cite{good60} and the following statement was made: ``{ \it the residual 
spin-flip amplitudes behave very much like spin-non-flip amplitudes at 
high energies and exhibit a pronounced forward peak which is largely 
independent of the particular elastic reaction chosen}''. 
This conclusion was reinforced by the analysis of similar $pp$ data 
\cite{hinotani79} when they became available. In this case, the 
situation is complicated by the existence of five independent helicity 
amplitudes in terms of which the polarization $P$ (also called transverse 
single-spin asymmetry \cite{buttimore99}) is defined as 
\beq
P=2{{\rm Im}[(\phi_1+\phi_2+\phi_3-\phi_4)\phi_5^{\ast}]\over 
[|\phi_1|^2+|\phi_2|^2+|\phi_3|^2+|\phi_4|^2+|\phi_5|^2]},
\eeq 
where $\phi_1,\phi_3$ are spin-non-flip amplitudes, $\phi_2,\phi_4$ 
are double spin-flip amplitudes and $\phi_5$ is a single spin-flip 
amplitude. 

In this work we are interested on high energy and not too high 
$t$ so that we can concentrate the analysis on the main aspects 
of the spin-flip amplitude for $pp$ scattering in the diffractive 
region. In \cite{buttimore99} it was analyzed the magnitude of 
$\phi_2$ and $\phi_4$ with respect to the spin-non-flip amplitudes and 
reasonable arguments were given concerning the linear dependence in 
$t$ for $\phi_2$ ($\phi_2\propto t$ when $t\to 0$). A similar 
dependence was also utilized in another work where the impact parameter 
space was used \cite{goloskokov91}. Neglecting $\phi_2$ and $\phi_4$ in
this work, we write \cite{hinotani79,neal69}
\beq
\phi_1+\phi_3\sim g(s,t),\quad \phi_5=h(s,t)
\label{effspamp}
\eeq
where $g(s,t)$ and $h(s,t)$ are {\it effective} spin-non-flip and 
spin-flip amplitudes, respectively. We then have
\cite{hinotani79,neal69}
\beq
P=2{{\rm Im}[g(s,t)h^{\ast}(s,t)]\over |g(s,t)|^2+|h(s,t)|^2} .
\eeq

Today, a rather considerable amount of data at higher energies has been 
gathered \cite{kline80} in the relatively small angle domain and new 
perspectives are being opened by the coming in operation of the 
Relativistic Heavy Ion Collider (RHIC), the ideal machine to study
polarization in high energy collision processes \cite{guryn00}.

In addition, our phenomenological information on the spin-non-flip 
amplitude is today much more complete and this can be used to reduce 
the uncertainties in the analysis. 

Using an explicit parametrization for the $pp$ spin-non-flip amplitude 
\cite{desgrolard00} whose parameters have been calculated against all 
high energy $pp$ and $\overline{p}p$ data (except polarization), we 
analyze the structure of the reduced spin-flip contribution. The following 
conclusions are reached analyzing the data:

\begin{description}
\item{a)} The (reduced) spin-flip amplitude has the typical peak in the 
forward direction which characterizes diffractive amplitudes or (which 
amounts for the same), has a non-negligible Pomeron $\IP$ contribution 
which
\item{b)} appears of comparable size as in the non-flip part\footnote{
The data \protect\cite{kline80} are analyzed introducing  
the {\bf effective} spin-flip amplitude reduced by explicitly factoring  
out the kinematical zero. For various reasons, related to the way one
removes the kinematical zero either by $\protect\sin\protect\theta$ or 
$\protect\sin(\protect\theta/2)$ or $\protect\sqrt{-t}$, it is 
difficult to compare our results for the fraction of the Pomeron $\IP$
contribution to the spin-flip amplitude with calculations from other 
works using different definitions for the single spin-flip amplitude. 
So, we adopt the notation of Refs. \protect\cite{predazzi67,hinotani79} 
and we do not compare the results with other approaches.} 
or not much smaller in the forward region, a characteristic already noticed 
at lower energies \cite{hinotani79};
\item{c)} the best way to fit the data is compatible with the same 
energy dependence in the spin-flip and in the spin-non-flip\footnote{
As we will see, a slower growth with energy of the spin-flip is not 
entirely ruled out but appears quite unlikely.}.
\item{d)} A zero of the polarization is predicted in the dip region; this 
zero recedes toward zero as the energy increases just as the dip
position does (in fact, it is the zero of the spin-non-flip amplitude 
that determines the zero of the polarization according to us);
\item{e)} the extrapolation to RHIC energies appears not very easy to 
measure.
\end{description}

\section{Definition of the amplitudes}

The $pp$ spin-non-flip amplitude is 
\beq
a_{pp}(s,t)=a_{+}(s,t)-a_{-}(s,t)
\eeq
with
\beq
a_{+}(s,t)=a_{\IP}(s,t)+a_f(s,t)\quad ,\quad
a_{-}(s,t)=a_{O}(s,t)+a_{\omega}(s,t)
\eeq
where $a_{P}(s,t)$ and $a_{O}(s,t)$ are the $\IP$ (Pomeron) and Odderon
amplitudes respectively and $a_f(s,t)$ [$a_{\omega}(s,t)$] are the 
even [odd] secondary Reggeons\footnote{Actually, $a_f$ embodies 
both $f$ and $\rho$ contributions (and $a_{\omega}$ both $\omega$ and 
$a_2$).}. These different amplitudes are taken directly from Ref. 
\cite{desgrolard00} and their explicit forms are listed in Appendix A 
together with the values of their parameters.

In the effective spin-flip amplitude, eq. (\ref{effspamp}), we neglect the 
contribution of secondary Reggeons and the simplest minded parametrization
is chosen\footnote{We have found that a much better result is obtained 
if the very small $|t|$ domain is singled out. Quite arbitrarily, we 
take $|t|\simeq 0.5\;{\rm GeV}^2$ as the limit for that region.}
\beqa
h(s,t)&=&a^{sf}(s,t) = (i\gamma_1+\delta_1)\sin\theta\;
\tilde{s}^{\alpha^{sf}(t)}e^{\beta_1 t}\Theta(|t|-0.5)   
\nonumber\\
&+& (i\gamma_2+\delta_2)\sin\theta\;
\tilde{s}^{\alpha^{sf}(t)}e^{\beta_2 t}\Theta(0.5-|t|),
\label{spinampl2}
\eeqa
where $\tilde{s}={s\over s_0}e^{-i\pi/2}$, $\Theta$ is the step function 
and we assume $s_0=1\;{\rm GeV^2}$ as in \cite{desgrolard00}.

To start with, we will take $\alpha^{sf}(t)$ to have exactly the same 
$\IP$ contribution of $a_{pp}$, {\it i.e.}
\beq
\alpha^{sf}(t)=\alpha_{\IP}(t)=\alpha_{\IP}(0)+\alpha_{\IP}'t
\label{alfsf}
\eeq
where $\alpha_{\IP}(0)$ and $\alpha_{\IP}'$ are found in Appendix A. The
data at $\sqrt{s}=13.8$, 16.8 and 23.8 GeV (a total of 64 points) are used in
the fit and the values of the parameters, together with the $\chi^2$ are 
listed in Table \ref{tab6param}.

\begin{table}[hbt!]
\centerline{
\begin{tabular}{|c|c|c|c|}
\hline
$\gamma_1$ & 2.55 & $\gamma_2$ & 0.18 \\
$\delta_1$ & 4.80 & $\delta_2$ & 0.45 \\
$\beta_1\;({\rm GeV}^{-2})$ & 6.25 & $\beta_2   
\;({\rm GeV}^{-2})$ & 2.30 \\ \hline
\multicolumn{4}{|c|}{$\chi^2/d.f.=1.1$}\\
\hline
\end{tabular}
}
\caption{Results from fitting polarization data at  
$\protect\sqrt{s}=13.8$, 16.8 and 23.8 GeV with 
eqs. (\protect\ref{spinampl2}) and (\protect\ref{alfsf}).}
\label{tab6param}
\end{table}

\begin{figure}[hbt!]
\begin{center}
\centerline{
\includegraphics[width=8cm,height=7.5cm]{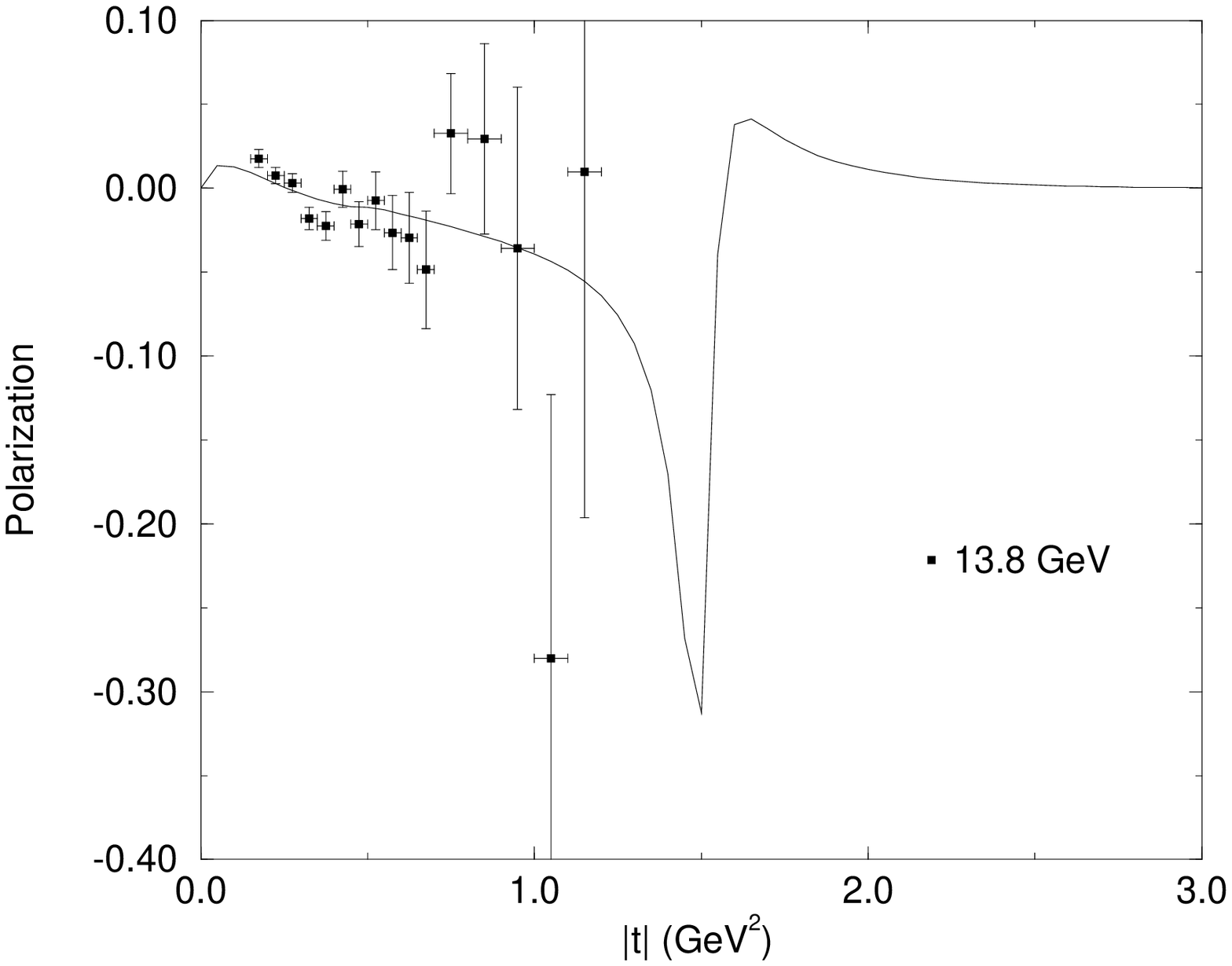}
\includegraphics[width=8cm,height=7.5cm]{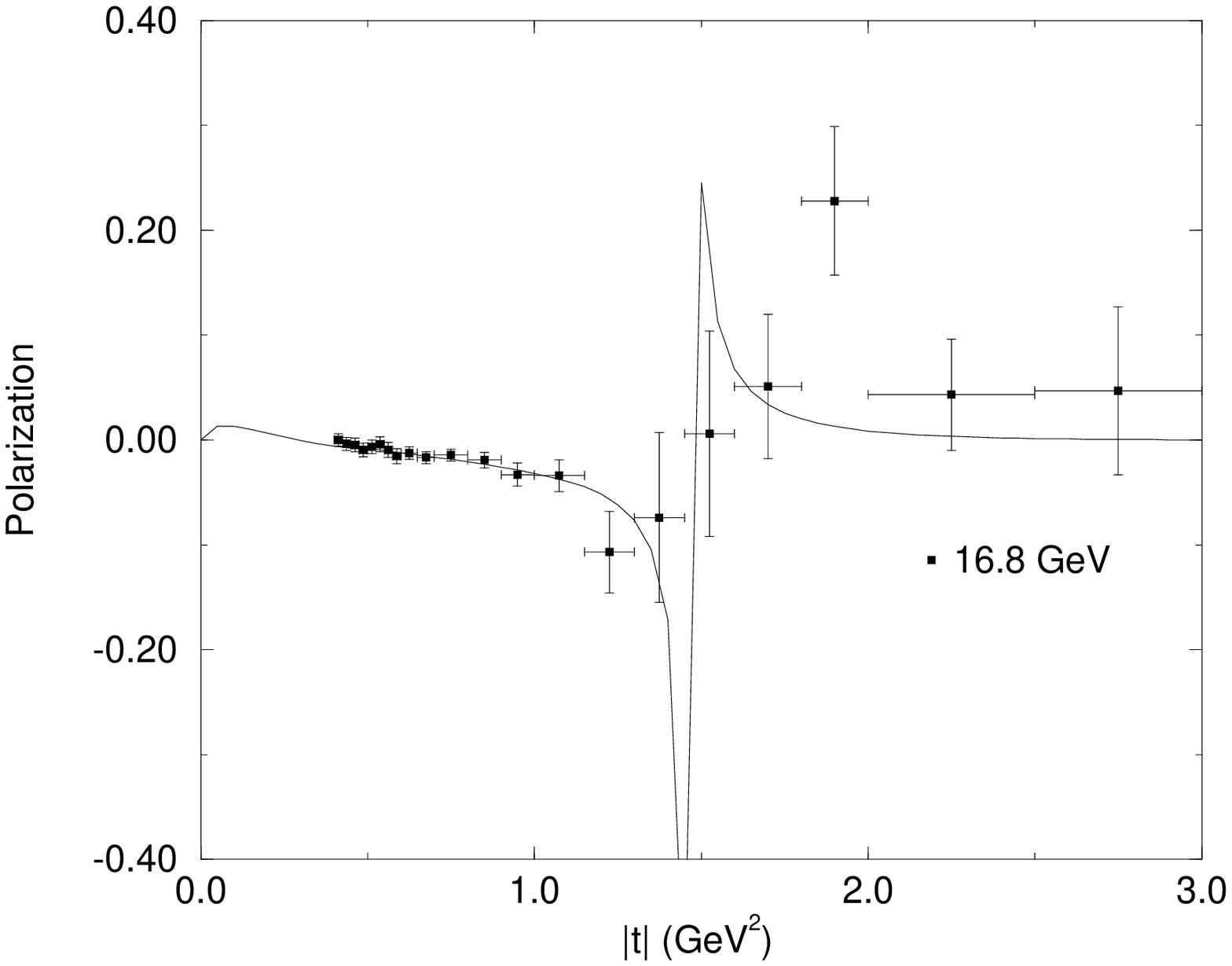}
}
\centerline{
\includegraphics[width=8cm,height=7.5cm]{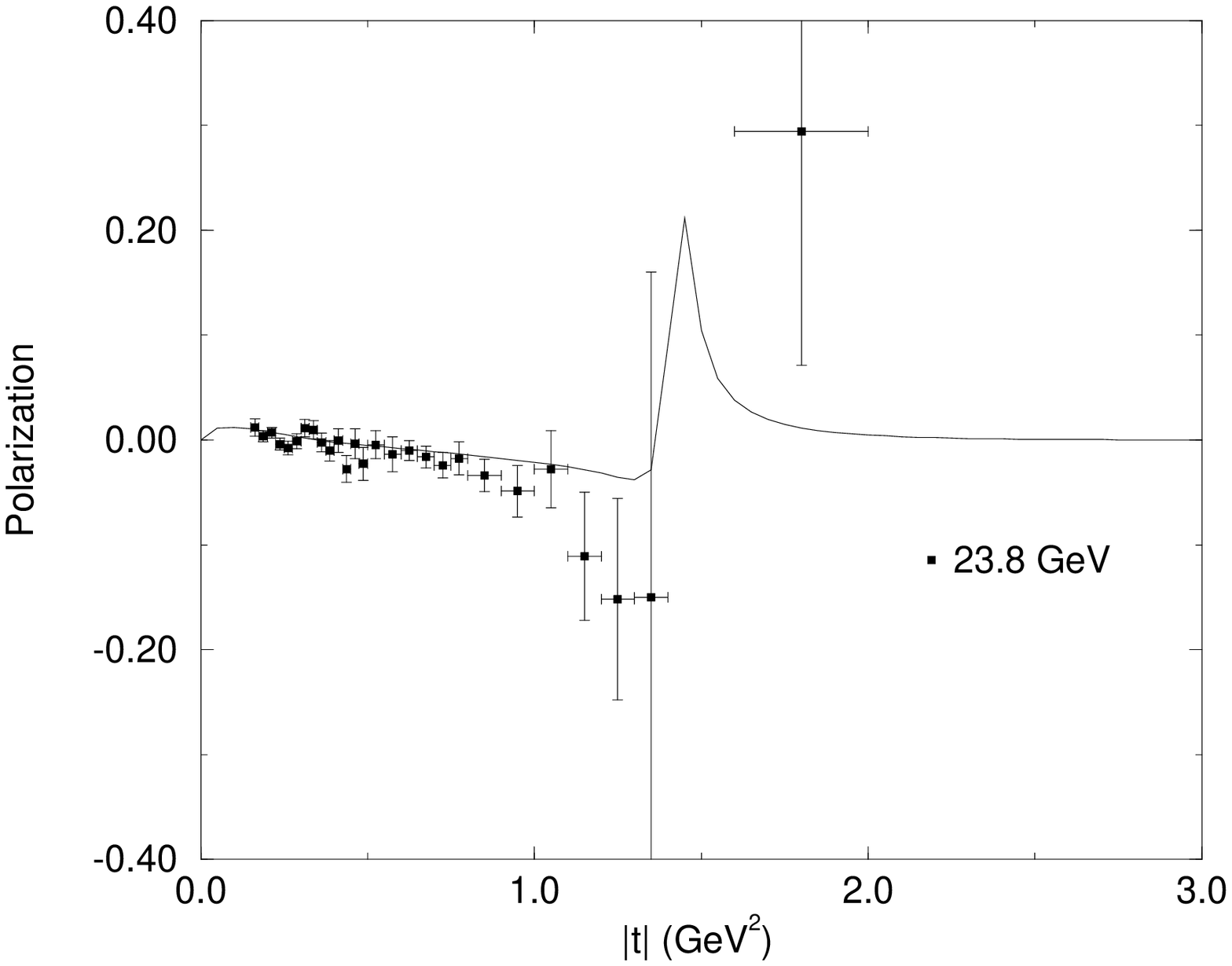}}
\caption{Results from fitting (see Table \protect\ref{tab6param}) 
at the energies of 13.8, 16.8 and 23.8 GeV.
}
\label{figpol6param}
\end{center}
\end{figure}

\begin{figure}[hbt!]
\begin{center} 
\centerline{
\includegraphics[width=8cm,height=7.5cm]{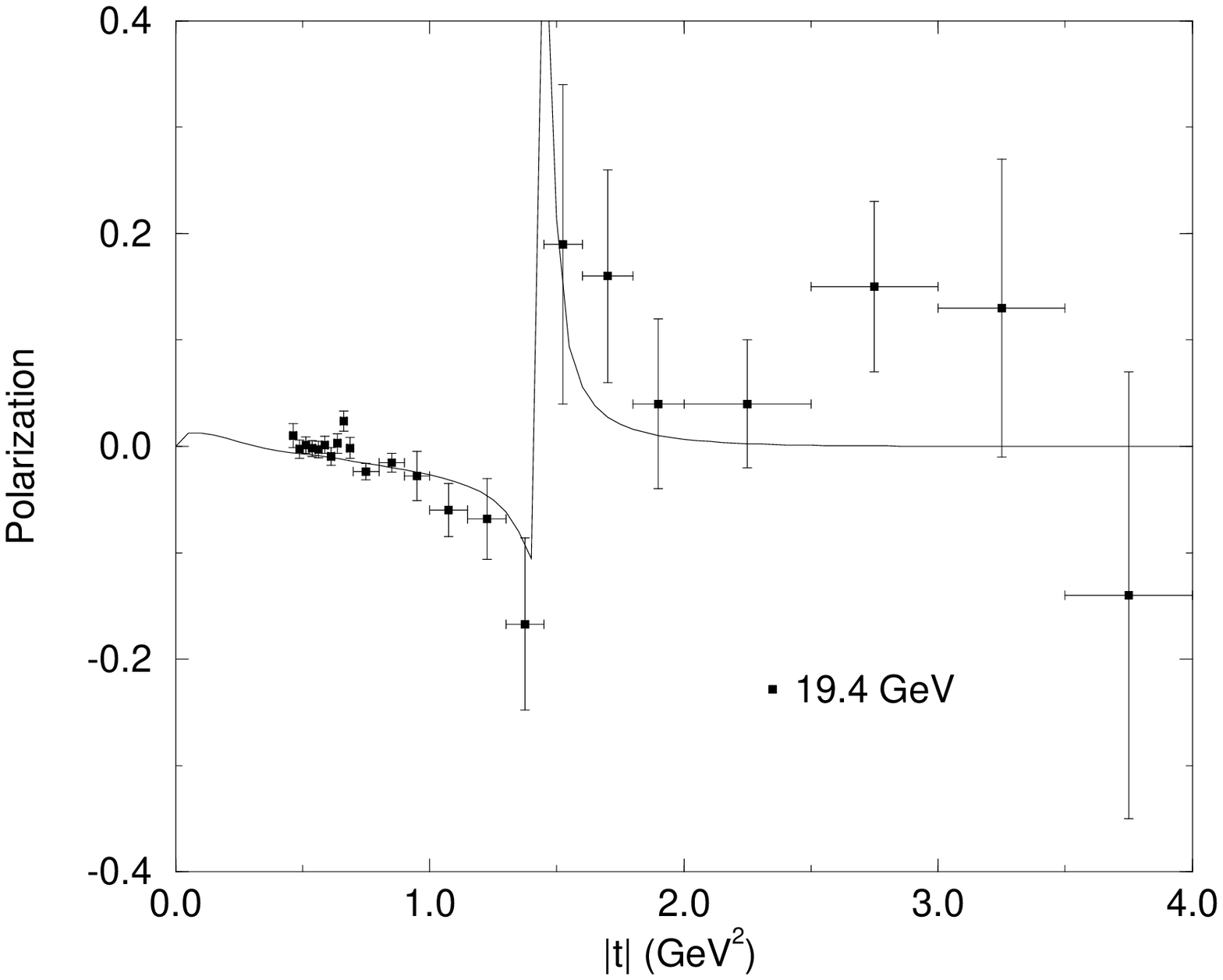}}
\caption{The prediction for polarization at 19.4 GeV (not used in the fit) 
compared with the experimental data on that energy.}
\label{figpol19pt4}
\end{center}
\end{figure}

In Fig. \ref{figpol6param} we show the polarization data together 
with our reconstruction. As a check of the validity of our solution, 
Fig. \ref{figpol19pt4} shows how it accounts for the data at 
$\sqrt{s}=19.4$ Gev (not used in the fit).


\begin{figure}[hbt!]
\begin{center}
\centerline{
\includegraphics[width=8cm,height=7.5cm]{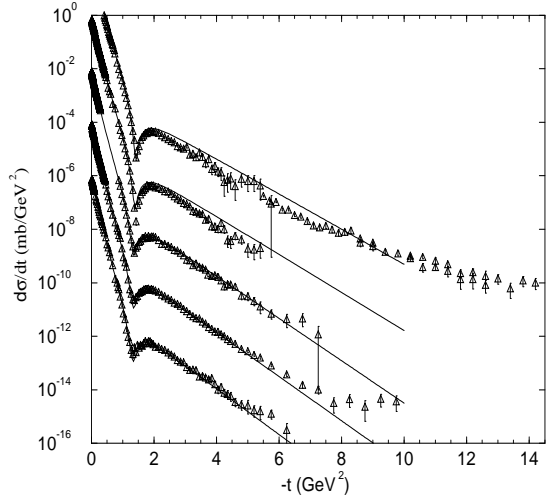}
}
\caption{The differential cross section obtained in this work taking 
into account the spin-flip amplitude. The highest set of data correspond 
to 23.5 and 27.4 GeV grouped together. The other sets (multiplied by 
powers of $10^{-2}$) are 30.5, 44.6, 52.8 and 62 GeV.
}
\label{figdsdt}
\end{center}
\end{figure}

\begin{figure}[hbt!]
\begin{center}
\centerline{
\includegraphics[width=8cm,height=7.5cm]{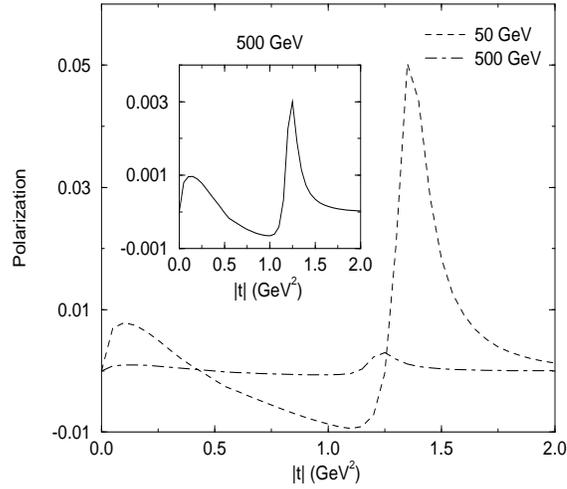}
}
\caption{The polarization predictions for 50 and 500 GeV 
with a detailed view of the 500 GeV (parameters from 
Table \protect\ref{tab6param}) in the inset.}
\label{figpol50500} 
\end{center}  
\end{figure}

Some considerations are in order:
\begin{description}
\item{a)} Our result, {\it i.e.} our effective spin-flip amplitude 
cannot be extended to $|t|$ values much higher than few ${\rm GeV}^2$ 
because the spin-non-flip amplitude utilized is at the Born level 
\cite{desgrolard00} and its description in the region after the 
dip ($|t|> 1.5\;{\rm GeV}^2$) is not very good; for this, it would be 
necessary to adopt the more sophisticated eikonalized version. Anyway, 
the $t$-region of interest for RHIC is up to $1.5\;{\rm GeV}^2$ 
\cite{guryn00} so we can concentrate the study on the not too high
$t$-region;
\item{b)} the solution we found shows a considerable $\IP$ contribution 
in the spin-flip amplitude. We will reconsider this point immediately 
below.
\item{c)} As a check that our introduction of the spin-flip amplitude 
has not spoiled the fit of Ref. \cite{desgrolard00}, we show 
$d\sigma/dt$ in Fig. \ref{figdsdt} at various energies.
\item{d)} The (small $|t|$) slope of the spin-flip amplitude 
$\beta_1=6.25\;{\rm GeV}^{-2}$ is somewhat different from the 
one that had been determined in Refs. \cite{hinotani79} but the 
present parametrization is considerably more elaborate and not directly 
comparable\footnote{The factor $\protect\tilde{s}^{\alpha(t)}$ 
generates a $|t|-$dependent factor $\exp(+\alpha't\ln s)$ which 
cooperates with $\exp(\beta_1t)$.}.
\item{e)} The extrapolation of our solution to 50 and 500 GeV 
predicts the polarization shown in Fig. \ref{figpol50500}. As one can 
see, it will be very hard to measure such a polarization. 
\end{description}

\begin{figure}[hbt!]
\begin{center}
\centerline{
\includegraphics[width=7cm,height=6.5cm]{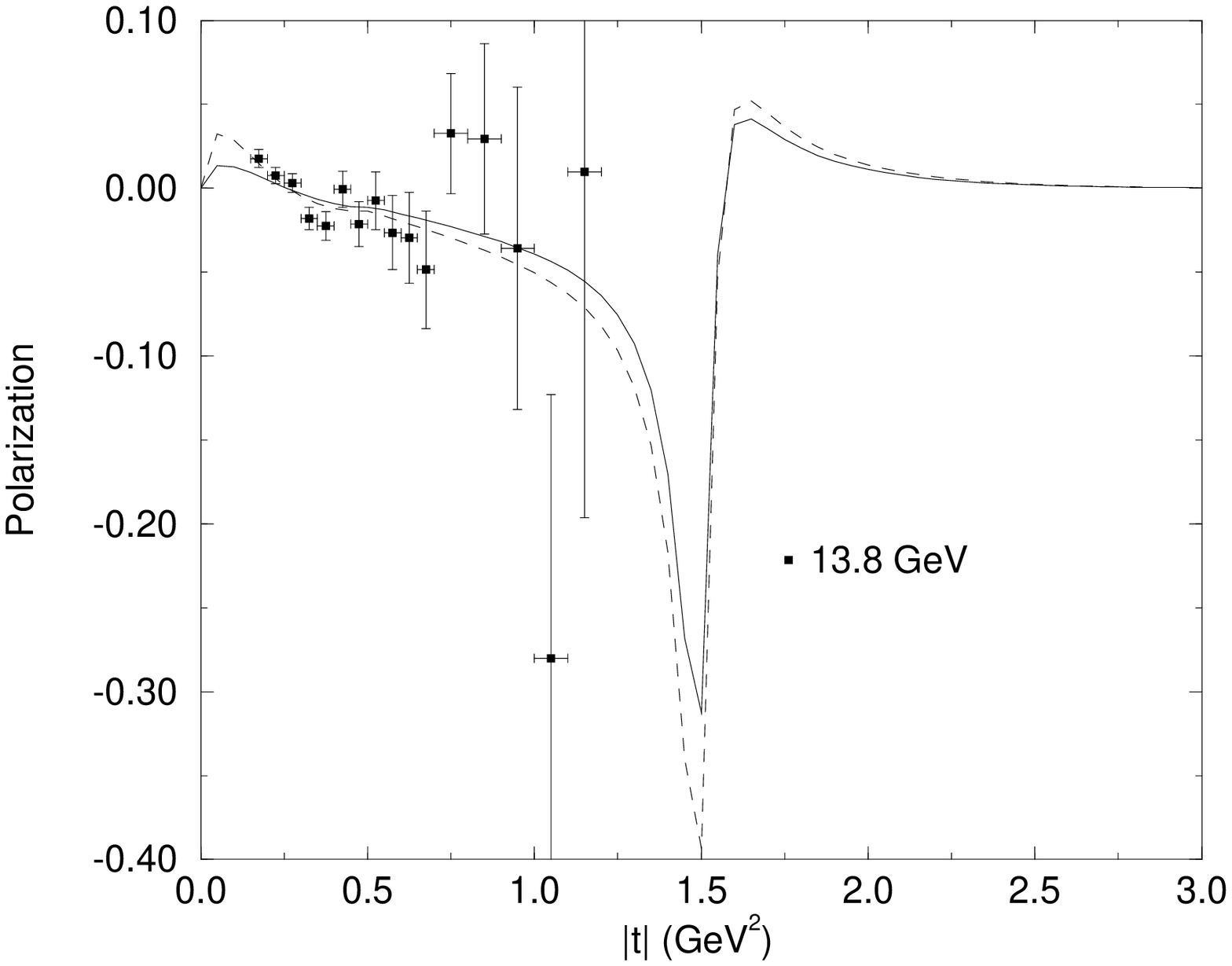}
\includegraphics[width=7cm,height=6.5cm]{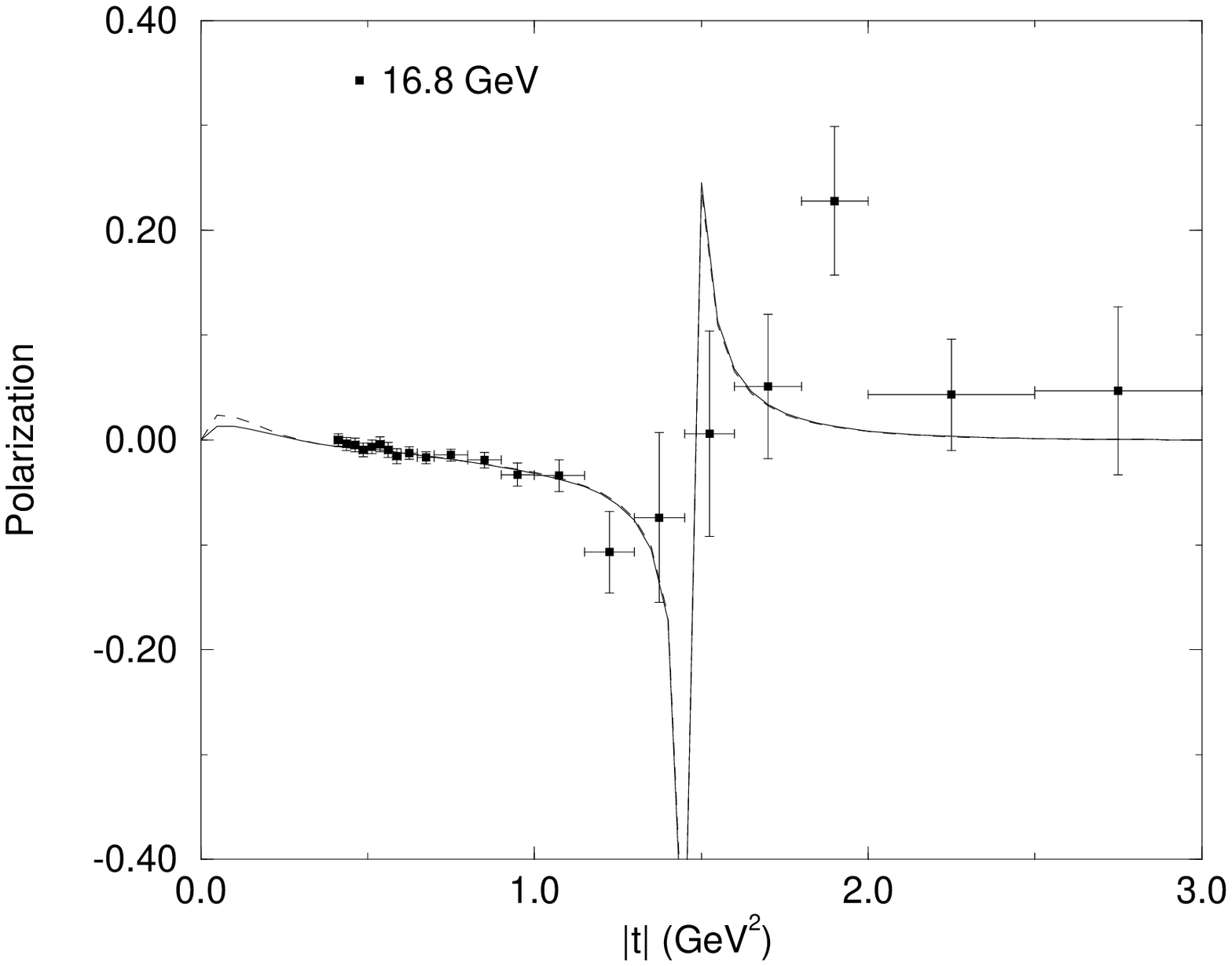}
}
\vspace{-0.5cm}
\centerline{
\includegraphics[width=7cm,height=6.5cm]{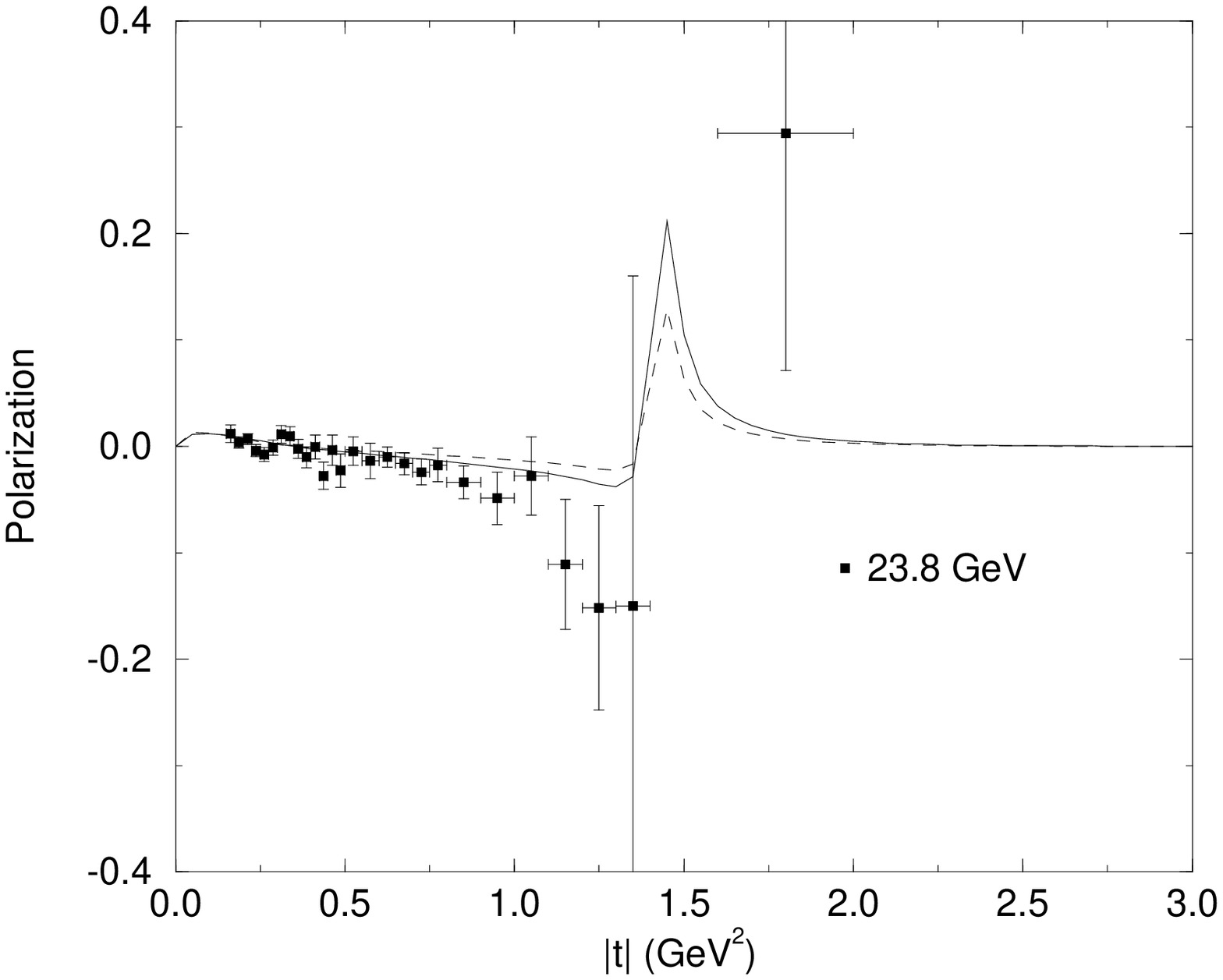}} 
\vspace{-0.5cm}
\caption{Results from fitting the data at 13.8, 16.8 and 23.8 GeV 
with values from Table \protect\ref{tab7param} (dashed line) compared 
to results from Table \protect\ref{tab6param} (solid line).
}
\label{figpol7param}
\end{center}
\end{figure}

\begin{figure}[hbt!]
\begin{center}
\centerline{
\includegraphics[width=8cm,height=7.5cm]{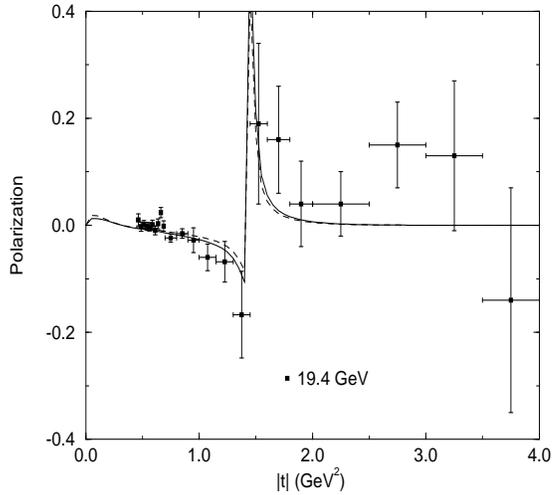}}
\caption{The prediction for polarization at 19.4 GeV compared with the
experimental data using parameters of 
Table \protect\ref{tab7param} (dashed line) and the result from 
Table \protect\ref{tab6param} (solid line).}
\label{figpol19pt4II}
\end{center}
\end{figure}

\begin{figure}[hbt!]
\begin{center}
\centerline{
\includegraphics[width=8cm,height=7.5cm]{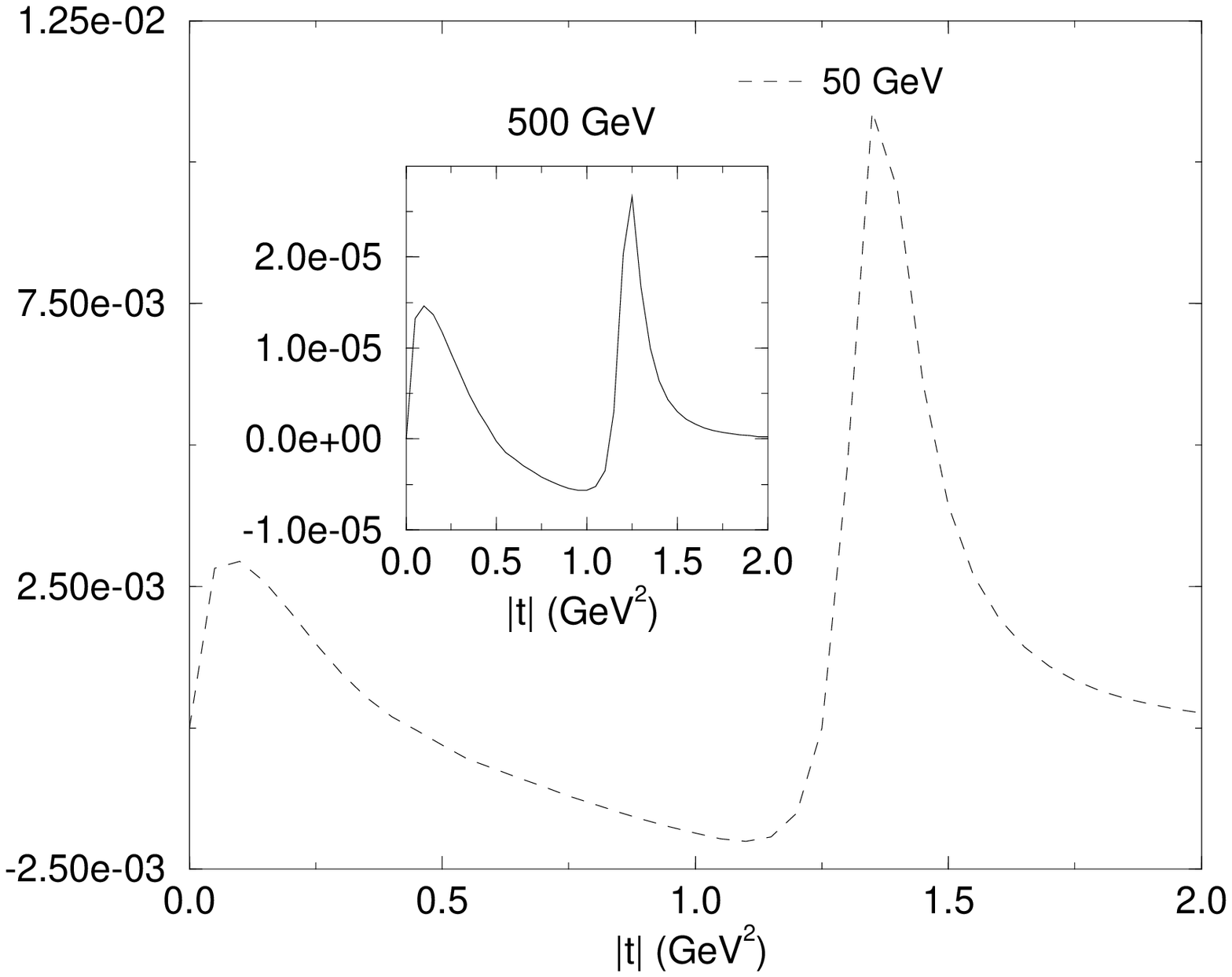}
}
\caption{The polarization predictions for 50 and 500 GeV using
the parameters of Table \protect\ref{tab7param}. A detailed view of the
500 GeV is shown in the inset.}
\label{figpol505007par}
\end{center}
\end{figure}

One open question remains the possibility that $\alpha^{sf}(0)$
be not the same as in the spin-non-flip amplitude. If $\alpha^{sf}(0)$
is left free to vary, one can still fit the data but the improvement on 
the description of the polarization is very small and visible only at 
small-$t$ for $\sqrt{s}=13.8$ GeV (see Fig. \ref{figpol7param}) while 
at other energies the differences in the description are irrelevant for 
the quality of the fitting (in fact, the $\chi^2/d.f.$ is practically 
the same in both cases) as can be seen in Fig. \ref{figpol7param} and 
\ref{figpol19pt4II}. Table \ref{tab7param} 
shows the values obtained for the parameters of eq. (\ref{spinampl2}) 
when $\alpha^{sf}(0)\neq\alpha_P(0)$. There appears a strong change of 
these parameters compensated by the new $\alpha^{sf}(0)$ and, as a 
result, the polarization is the same up to $\sqrt{s}=23.8$ GeV but when 
we look at RHIC energies the polarization predictions are even smaller
(Fig. \ref{figpol505007par}), for example at $\sqrt{s}=500$ GeV the 
result is one hundred times smaller making essentially impossible to 
detect these values experimentally. At this point it is hard to conclude 
if those new values (Table \ref{tab7param}) are an improvement of the
fitting or another local minimum. Since the larger number of fitted 
parameters did not improve the description of polarization 
data we conclude that a slower growth with energy of the spin-flip 
amplitude is quite unlikely as already mentioned. We do not show 
$d\sigma/dt$ calculated with the values of Table \ref{tab7param} 
because it is indistinguishable from Fig. \ref{figdsdt}.

\begin{table}[hbt!]
\centerline{
\begin{tabular}{|c|c|c|c|}
\hline
\multicolumn{4}{|c|}{$\alpha_{P}^{sf}(0)=0.377$} \\
\hline
$\gamma_1$ & -300 & $\gamma_2$ & -16 \\
$\delta_1$ & 443 & $\delta_2$ & 19 \\
$\beta_1\;({\rm GeV}^{-2})$ & 7.84 & $\beta_2
\;({\rm GeV}^{-2})$ & 2.32 \\ \hline
\multicolumn{4}{|c|}{$\chi^2/d.f.=1.1$}\\
\hline
\end{tabular}
}
\caption{Results for eq. (\protect\ref{spinampl2}) with
$\protect\alpha^{sf}(0)\protect\neq\protect\alpha_{P}(0)$.}
\label{tab7param}
\end{table}

\section{Conclusions}

The conclusions have already been anticipated 
but, once more, spin effects appear extremely interesting:
the spin-flip amplitude confirms the diffractive-like behavior that
had been anticipated by the pioneer analysis of long ago 
\cite{predazzi67,hinotani79}. Analyzing the reduced
spin-flip amplitude (removing the kinematical zero by the $\sin\theta$ 
factor in eq. (\ref{spinampl2})), we find that it is comparable in size 
with the spin-non-flip amplitude as already noticed in \cite{hinotani79}. 
The spin-flip amplitude can be represented with a very simple form 
and the general features of the polarization data are 
well described if the same energy dependence in spin-flip as in 
spin-non-flip Pomeron amplitude is utilized. The 
spin-flip amplitude appears to have two different regimes, a fast 
decrease with $|t|$ at small-$t$ with $\beta_1=6.25\;{\rm GeV}^{-2}$
followed by a slower decrease at medium-$t$ 
($\beta_1=2.30\;{\rm GeV}^{-2}$). 
We calculated the polarization at RHIC energies and predicted two 
zeros, the last one is determined by the zero of the spin-non-flip 
amplitude followed by a local maximum. However, the 
magnitude of the polarization becomes so small when the energy increases 
that it may be very difficult to perform the experimental 
measurements at $\sqrt{s}\sim 500$ GeV.

It would be possible to improve the description of the data with more 
elaborated forms for spin-non-flip (by eikonalizing it
\cite{desgrolard00}, for example) and spin-flip (introducing
non-asymptotic effects by secondary Reggeons) amplitudes but the main 
aspects of our analysis would be the same for small-$t$ (where the Born 
amplitudes work well) and higher energies (the region of Pomeron 
dominance). 

{\bf Acknowledgements}
One of us (AFM) would like to thank the Department of Theoretical Physics
of the University of Torino for its hospitality and the FAPESP of Brazil
for its financial support. Several discussions with Prof. E. Martynov and 
Prof. M. Giffon are gratefully acknowledged.

\vspace{1.0cm}
\centerline{\bf APPENDIX}
\appendix
\section{The spin-non-flip amplitude}

The spin-non-flip amplitude utilized in this work is
\beq
a_{pp}(s,t)=a_{+}(s,t)-a_{-}(s,t) ,
\eeq
where

\beq
a_{+}(s,t)=a_{\IP}(s,t)+a_f(s,t) \; {\rm and} \;
a_{-}(s,t)=a_{O}(s,t)+a_{\omega}(s,t).
\eeq

The expressions for the two Reggeons used in \cite{desgrolard00} are

\beq
a_R(s,t)=a_R\tilde{s}^{\alpha_R(t)}e^{b_R t},\;
\alpha_R(t)=\alpha_R(0)+\alpha_R't,\; (R=f\; {\rm and} \; \omega)
\eeq
with $a_f (a_{\omega})$ real (imaginary). 

For the Pomeron, the non spin-flip amplitude is
\beq
a_{\IP}^{(D)}(s,t)=a_{\IP}\tilde{s}^{\alpha_{\IP}(t)}
[e^{b_{\IP}(\alpha_{\IP}(t)-1)}(b_{\IP}+ln\tilde{s})+d_{\IP}ln\tilde{s}] 
\eeq
while for the Odderon, we choose

\beq
a_{O}(s,t)=(1-\exp(\gamma t))*a_O\tilde{s}^{\alpha_O(t)}
[e^{b_O(\alpha_O(t)-1)}(b_O+ln\tilde{s})+d_Oln\tilde{s}],
\eeq
and again $a_{\IP} (a_O)$ real (imaginary). We use 
$\alpha_i(t)=\alpha_i(0)+\alpha_i't$ where $i=\IP ,O$.

Our definition for the amplitude follows \cite{desgrolard00} so 
that
\beqa
\sigma_t={4\pi\over s}{\rm Im}\{ a_{pp}(s,t=0)\},
\label{sctot} \\
{d\sigma\over dt}={\pi\over s^2}(|a_{pp}(s,t)|^2+|a^{sf}(s,t)|^2).
\label{scdif}
\eeqa

In this work we retain the same parameters for the spin-non-flip
amplitude as in \cite{desgrolard00} and we keep them fixed while 
fitting the parameters of the spin-flip
amplitude. We utilize the dipole model at the Born level since great 
part of the polarization data is contained in the $t$-domain 
corresponding to the region before the dip in $d\sigma/dt$ (well 
described without eikonalization). The values of the parameters of 
the spin-non-flip amplitude are shown in Table \ref{tabsnfpar}.

\begin{table}[hbt!]
\centerline{
\begin{tabular}{|c|c|c|c|c|}
\hline
 & Pomeron & Odderon & $f$-Reggeon & $\omega$-Reggeon \\ 
\hline
$\alpha_i(0)$ & 1.071 & 1.0 & 0.72 & 0.46 \\
$\alpha_i'$ & 0.28 ${\rm GeV}^{-2}$ & 0.12 ${\rm GeV}^{-2}$ & 
0.50 ${\rm GeV}^{-2}$ & 0.50 ${\rm GeV}^{-2}$ \\
$a_i$ & -0.066 & 0.100 & -14.0 & 9.0 \\
$b_i$ & 14.56 & 28.10 & 1.64 ${\rm GeV}^{-2}$ & 0.38 ${\rm GeV}^{-2}$ \\
$d_i$ & 0.07 & -0.06 & - & - \\ 
$\gamma$ & - & 1.56 ${\rm GeV}^{-2}$ & - & - \\
\hline
\end{tabular}
}
\caption{Parameters of the dipole model at the Born level
\protect\cite{desgrolard00} with $i=\IP ,O,f,\omega$.}
\label{tabsnfpar}
\end{table} 

To calculate the polarization we utilized the form
\beq
P=2
{{\rm Im}(a_{pp}(s,t)(a^{sf}(s,t))^{\star})\over 
|a_{pp}(s,t)|^2+|a^{sf}(s,t)|^2} ;
\eeq
where the star on the numerator means the complex conjugate.


\begin{thebibliography}{99}
\bibitem{predazzi67}E. Predazzi and G. Soliani, Nuovo Cimento {\bf 51}A,
427 (1967).
\bibitem{good60}M.L. Good and W.D. Walker, Phys. Rev. {\bf 120} 1857
(1960).
\bibitem{hinotani79}K. Hinotani, et al., Nuovo Cim. {\bf 52}A 363 (1979).
\bibitem{buttimore99}N.H. Buttimore et al., Phys. Rev. D{\bf 59} 114010
(1999).
\bibitem{goloskokov91}S.V. Goloskokov et al., Z. Phys. C{\bf 50} 455 
(1991).
\bibitem{neal69}H.A. Neal and E. Predazzi, Nuovo Cim. {\bf 62}A 275
(1969).
\bibitem{kline80}R.V. Kline et al., Phys. Rev. D{\bf 22} 553 (1980); 
G. Fidecaro et al., Nucl. Phys. B{\bf 173} 513 (1980); G. Fidecaro et al., 
Phys. Lett. B{\bf 105} 309 (1981); J. Snyder et al., Phys. Rev. Lett. 
{\bf 41} 781 (1978).
\bibitem{guryn00}W. Guryn, Talk given at the Diffraction 2000, September
2-7, 2000, Cetraro, Italy.
\bibitem{desgrolard00}P. Desgrolard et al., Eur. Phys. J. C{\bf 16} 499 
(2000).
\end{thebibliography}
\end{document}